# Tunable cavity coupling to spin defects in 4H-silicon-carbide-on-insulator platform

Tongyuan Bao[1,†], Qi Luo[1†], Ailun Yin[3,5†], Yao Zhang[1], Haibo Hu[1,2], Zhengtong Liu[2], Shumin Xiao[1,2,6], Xin Ou[3,5*], Yu Zhou[1,4*], Qinghai Song[1,2,4,6*]

1. Ministry of Industry and Information Technology Key Lab of Micro-Nano Optoelectronic Information System, Guangdong Provincial Key Laboratory of Semiconductor Optoelectronic Materials and Intelligent Photonic Systems, Harbin Institute of Technology, Shenzhen 518055, P. R. China.
2. Pengcheng Laboratory, Shenzhen 518055, P. R. China.
3. State Key Laboratory of Functional Materials for Informatics, Shanghai Institute of Microsystem and Information Technology, Chinese Academy of Sciences, Shanghai 200050, China
4. Quantum Science Center of Guangdong-HongKong-Macao Greater Bay Area (Guangdong), Shenzhen 518045, China
5. The Center of Materials Science and Optoelectronics Engineering, University of Chinese Academy of Sciences, Beijing 100049, China.
6. Collaborative Innovation Center of Extreme Optics, Shanxi University, Taiyuan 030006, Shanxi, P. R. China.

† These authors contributed equally to this work.
*Corresponding authors: zhouyu2022@hit.edu.cn, ouxin@mail.sim.ac.cn or qinghai.song@hit.edu.cn

**Abstract:**

**Silicon carbide (SiC) has attracted significant attention as a promising quantum material due to its ability to host long-lived, optically addressable color centers with solid-state photonic interfaces. The CMOS compatibility of 4H-SiCOI (silicon-carbide-on-insulator) makes it an ideal platform for integrated quantum photonic devices and circuits. While micro-ring cavities have been extensively studied in SiC and other materials, the integration of 4H-SiC spin defects into these critical structures, along with continuous mode tunability, remains unexplored. In this work, we demonstrate the integration of PL4 divacancy spin defects into tunable micro-ring cavities in scalable thin-film 4H-SiC nanophotonics. Comparing on- and off-resonance conditions, we observed an enhancement of the Purcell factor by approximately 5.0. This enhancement effectively confined coherent photons within the coupled waveguide, leading to a twofold increase in the ODMR (optically detected magnetic resonance) contrast and coherent control of PL4 spins. These advancements lay the foundation for developing SiC-based quantum photonic circuits.**

**Main:**

Long-lived and optically addressable spin defects in silicon carbide [1,2] are emerging as a promising quantum candidate due to their robustness, versatile control [3–6], and developed read-out methods [7,8]. One advantage of silicon carbide compared to diamond is its compatibility with complementary metal-oxide-semiconductor (CMOS) fabrication technology [9,10]. This makes 4H-SiCOI (silicon-carbide-on-insulator) a promising platform for advancing integrated quantum photonic devices and circuits [9–13]. However, leveraging 4H-SiCOI for the continued scaling up quantum networks presents a significant challenge: achieving precise and continuous tunability of the interaction between photonic cavities and spin defects [12,14–17]. This is especially important for enhancing zero-phonon line (ZPL) emissions over those in the phonon sideband. Boosting emissions into the ZPL is essential for achieving key milestones in the construction of quantum networks, such as single-shot readout [7,18,19], spin-photon entanglement [20,21], and heralded between remote nodes [22–24], where coherent and indistinguishable photons are needed.

Although cavity-enhanced color centers have been demonstrated in 4H-SiC, the majority of those implementations use 1D nanobeam or 2D photonic crystal cavities with bulk carving [14,15,25,26] and undercut techniques [27], limiting functionality regarding photonic intersections. Achieving quantum computation and error-protected communication ultimately requires developing highly complex photonic circuits that leverage the integrated photonics expertise accumulated over the past two decades [9,28]. With the advancement of quantum-grade 4H-SiC thin film techniques [12], another approach is to develop spin-coupled and tunable 2D cavities like micro-ring cavities with in-plane photonic waveguides for the confinement of those coherent photons. Micro-ring cavities are vital for large-scale integrated photonic networks on a chip, particularly for applications involving on-chip quantum sources [29], nonlinear frequency conversion [12], and filtering. While micro-ring cavities have been farbricated in SiC [30,31] and other materials [29,32], the integration of 4H-SiC spin defects into this critical structure, along with continuous mode tunability, has not yet been achieved. In this work, we demonstrate the integration of PL4 divacancy spin defects into tunable micro-ring cavities in thin-film 4H-SiC. This lays the foundation for the development of SiC-based quantum photonic circuits.

**PL4 divacancy spins on 4H-SiCOI**

Silicon carbide (SiC) divacancies are specific point defects in the SiC crystal lattice, formed by the absence of a silicon atom and an adjacent carbon atom [33,34]. These divacancies exhibit excellent optical and spin properties, making them highly promising for quantum technology applications [35]. The labels PL1-PL7 correspond to distinct defect types associated with various configurations of these divacancies in the 4H-SiC polytype [34,36]. Among these, the PL4 defect is particularly interesting, as it frequently appears as the most prominent peak in divacancy-related photoluminescence (PL) spectra generated via carbon ion implantation [37]. This prominence suggests that the PL4 configuration is more readily formed in the fabrication.

To mitigate the adverse effects of ion irradiation on the spin properties, the SiCOI wafer employs a thinning and polishing technique [12] rather than the conventional smart cut method [10]. As illustrated in Figure 1a, a 4H-SiC wafer with an epitaxial and thin $SiO_2$ layer is bonded to an oxidized Si wafer. The bonded SiC layer undergoes mechanical grinding followed by chemical-mechanical polishing (CMP) to achieve a thickness of several micrometers. A SiCOI wafer with a designed thickness of 200 nm is obtained after dry etching the SiC layer. Divacancy spins are generated in the SiC membrane via carbon ion implantation. The implantation was performed using 30 keV ions at a dose of $1.0\times10^{13}$ cm$^{-2}$. This was followed by annealing in a high vacuum at 900°C for 30 minutes to repair any remaining lattice damage. The fabrication of photonic devices began with the deposition of a 30 nm Cr layer on the SiCOI membrane. The micro-ring cavities with different diameters were then patterned onto an electron beam resist (ZEP 520A) using an electron beam lithography overlay technique. After developing the resist, the Cr mask was created through ICP dry etching. The photonic devices were fabricated by etching the SiC membrane with the patterned Cr as a hard mask.

After the Cr lift-off, the optical image of the fabricated micro-ring cavities with diameters ranging from 7.3 $\mu m$ to 8.5 $\mu m$ are displayed in Figure 1b (refer to Figure S4 for the device design and simulation details). The top-view scanning electron microscope (SEM) images of the micro-ring cavities with a diameter of 8.1 $\mu m$ is displayed in Figure 1c. The fabricated device's dimensions closely align with our design specifications. As depicted in Figure 1d, the fabricated micro-ring cavities were characterized using a home-built confocal microscope system operating at 4K. A 914 nm diode laser was used to excite the vacancy spins. The fluorescence was then collected and detected with a superconducting nanowire single photon detector (SNSPD). Based on the total emission rate, we estimate the concentration of the PL4 centers to be $5\times10^{14}$ centers/cm². The emission is guided to a 90:10 fiber beam splitter either for spectrum and lifetime measurements (90%, Channel 1) or for ODMR measurements (10%, Channel 2), respectively. A blazed grating (600/mm) with a rotational state was utilized in the PL measurement and ZPL filtering. The dependence of the angle and wavelength was pre-characterized by a tunable diode laser with a wavemeter.

**Optical characterization of the micro-ring cavities at 4K**

Optical characterization at 4K reveals detailed insights into the device's performance, as illustrated in Figure 2. Figure 2a displays the confocal scan map of the micro-ring cavity, highlighting two coupling waveguides and four grating couplers that align with the designed geometry. By maintaining excitation within the micro-ring and adjusting the collection angle using a scanning mirror, we captured the angle-dependent scan map, as depicted in Figure 2b. This scan map clearly shows four emission spots from each grating coupler, confirming the successful coupling of divacancy photoluminescence (PL) from the ring to the waveguide and side grating couplers. In addition to excitation at the ring, we also conducted PL spectrum measurements through excitation at the waveguide, as detailed in Figure S3. To confirm the effective coupling between the spin defects and the microcavity, we measured the PL spectrum collected from both the confocal spot and grating 1. The full spectrum of divacancy color centers under 0.2 mW confocal excitation is depicted in Figure 2c. The spectrum is dominated by the zero-phonon line (ZPL) from PL4, with additional ZPLs from PL1, PL2, and PL6 at much lower intensities [34]. The PL spectrum in Figure 2d was obtained from grating 1 under specific power excitation at the side of the micro-ring. The spectra clearly show multiple cavity modes within the emission band of the PL4 divacancy spin. Different colors represent micro-ring cavities with varying diameters ranging from 7.3 μm to 8.9 μm, with free-spectral range (FSR) from $14.8\pm0.07$ nm to $17.2\pm0.02$ nm. As the diameter changes, the shifts in the optical modes can be observed indicated by the dashed line. This shift demonstrates the coarse tunability (13 nm in total) of optical modes by altering the ring's pre-designed dimensions. However, the zero-phonon line (ZPL) from PL4 is weak and also marked with a dashed line, indicating it is not effectively coupled with the existing modes. As displayed in Figure 2e, the measured cavity mode coupled with PL4 PL shows a quality factor Q of $1261\pm39$. The quality factors of the other microring resonators with diameters of 7.3 μm, 7.7 μm, 8.5 μm, and 8.9 μm were measured to be $1188\pm55$, $1218\pm72$, $1181\pm88$, and $943\pm33$, respectively. The relatively low Q factor is likely due to surface roughness, a consequence of imperfections in the etching process, which is visible in the rough sidewalls of the micro-ring (as shown in Figure S8). These rough surface conditions cause substantial scattering and absorption losses as light propagates through the cavity [38]. Thus, the enhancement can be further optimized by optimizing the etching parameters. Besides characterization at 4K, we also conducted room-temperature measurements of the fabricated micro-ring cavities, summarized in Figure S1. Similar cavity modes can also been observed.

**Continuous cavity mode tunning via gas condensation**

After conducting the initial optical characterization and confirming the coupling between the spin defects and the cavity, the tuning of the cavity mode and the strong enhancement of the zero-phonon line (ZPL) of PL4 within the cavity were demonstrated. A cavity with an 8.1 μm diameter was selected, whose optical mode is slightly blue-shifted compared to the ZPL wavelength. The optical mode can be gradually redshifted using nitrogen gas condensation [39]. Figure 3a displays an intensity map showing the interaction between the cavity mode and the ZPL of PL4 as a function of gas injection steps. The cavity mode undergoes red shifting while the ZPL wavelength remains constant. A significant enhancement in emission is observed when the two intersect on the map. To facilitate the explanation of subsequent lifetime and PL spectrum measurements, seven points (A-F) are marked throughout the process. During the first three cycles, approximately 0.05 L of nitrogen gas at a pressure of 100 Pa was introduced into the cryostat per cycle. From that point until point F, the same volume of nitrogen gas (0.05 L) was introduced per cycle, but at a reduced pressure of 15 Pa. After point F, the introduction pressure was adjusted to 50 Pa while maintaining the same volume per cycle. This tuning process is reversible, and the system can return to its pre-tuned state simply by increasing the temperature. To further demonstrate the versatility of this tuning method, we applied it to modulate the cavity modes of another microring resonator with a diameter of 7.7 μm, as shown in Fig. S6.

Figure 3b illustrates the PL spectrum at Point A (off-resonance) and Point D (on-resonance). The ZPL intensity experiences a 36-fold increase, with other modes showing no significant variation. This enhancement occurs only at the output coupler, as evidenced by the two similar confocal PL spectra measurements between on- and off-resonance shown in Figure 3c. This suggest that the enhanced coherent photons were effectively confined within the coupled waveguide. Lifetime measurements were performed using a 940 nm picosecond laser with a repetition rate of 10 MHz. As shown in Figure 3d, the temporal profile of the PL emission was fitted by a single exponential function, indicating a reduction in the lifetime from $15.85 \pm 0.09$ ns in the off-resonance case to $13.64 \pm 0.07$ ns when resonant with the cavity mode.

Next, we analyze the Purcell effect observed in our experiment. When the cavity mode is off-resonance, the total emission rate of uncoupled color centers is given by: [17]

$$\frac{1}{\tau_{\text{off}}} = \frac{1}{\tau_{\text{ZPL}}} + \frac{1}{\tau_{\text{PSB}}},$$

where $1/\tau_{\text{ZPL}}$ and $1/\tau_{\text{PSB}}$ are the emission rates into ZPL and phonon sideband (PSB), respectively. When the cavity mode is on-resonance, the emission rate is modified as: [17]

$$\frac{1}{\tau_{\text{on}}} = \frac{F+1}{\tau_{\text{ZPL}}} + \frac{1}{\tau_{\text{PSB}}},$$

where F is the Purcell factor, which can be estimated from the measured lifetimes using the relation:

$$F = \frac{1}{\xi_{\text{ZPL}}} \left( \frac{\tau_{\text{off}}}{\tau_{\text{on}}} - 1 \right),$$

Where the ratio $\xi_{\text{ZPL}} = \tau_{\text{off}}/\tau_{\text{ZPL}}$ is equivalent to the branching ratio into the ZPL (Debye-Waller factor, DWF). The DWF was measured to be 0.031 based on the integration of the measured PL spectrum in Figure 2c. This DWF is consistent with the theoretically calculated value of 0.038 [40], corresponding to a Purcell factor of 5.23.

In the above calculation, we assume that $1/\tau_{\text{off}}$ is equal to the lifetime of centers in the un-patterned SiC membrane $1/\tau_0$. However, due to the presence of the photonic band gap, the optical density of the state is reduced, typically resulting in $\tau_{\text{off}}$ being generally slightly greater than $\tau_0$ [15,41]. The Purcell factor can also be estimated using the following relation [15]

$$F = \frac{\tau_0}{\text{DWF}} \left( \frac{1}{\tau_{\text{on}}} - \frac{1}{\tau_{\text{off}}} \right),$$

where DWF is Debye–Waller factor. $\tau_0$ is measured to be $14.94 \pm 0.08$ ns (Fig. S2 in the Supplementary Note), corresponding to a Purcell factor 4.9. The lifetimes of various points of the whole gas condensation process from Figure 3a are summarized in Figure 3f. As anticipated, the closer the cavity mode is to the ZPL position, the more significant the reduction in lifetime.

**ODMR and coherent control of the PL4 spins in micro-ring cavities**

Generating specific defect types among ensemble divacancy spins via implantation is extremely challenging because all polytopes may form. In on-chip sensing, the on-resonance enhancement of the specific ZPL serves as a filtering mechanism and increases the ODMR signal amplitude [26]. To perform spin-dependent measurements on defects within the cavity structure, a 20 μm copper wire is used as a microwave antenna near the device in the cryostation. The PL4 defect is oriented along the basal planes, resulting in lower $C_{1h}$ symmetry and non-degenerate spin transitions at zero magnetic fields (B=0). Figure 4a presents the zero-field ODMR measurements obtained from both confocal detection and grating 1. The spectrum reveals two prominent peaks at 1315.1±0.3MHz and 1352.4±0.2MHz, which align with the reported values of 1.316 GHz and 1.353 GHz in the literature [34]. The ODMR contrast improves by a factor of two when the PL emission is collected via the grating (on resonance) instead of confocally (off-resonance). This enhancement demonstrates that the micro-ring cavity functions as a filter, increasing the proportion of PL4 photons in the total collected emission. Finally, the coherent control of PL4 spins both with and without coupling to the micro-ring cavity has been performed. Figure 4b shows the Rabi oscillations of the PL4 divacancy spin, the contrast of Rabi rotation is consistent with the ODMR measurements.

**Conclusion**

In conclusion, we have successfully demonstrated the integration and control of spin defects in tunable micro-ring cavities within thin-film 4H-SiC nanophotonics. By leveraging the cavity's resonance enhancement, a substantial 36-fold increase in zero-phonon line (ZPL) intensity was achieved, along with a twofold increase in optically detected magnetic resonance (ODMR) contrast compared to the off-resonance conditions. Coherent control of the micro-ring cavity integrated PL4 spins was realized, highlighting the potential for integrating such systems into more complex quantum photonic circuits. Our work provides important insights into defect-cavity interactions, paving the way for developing advanced silicon carbide-based quantum photonic devices with enhanced functionalities.

**Methods**
**4H-SiCOI and micro-ring cavity fabrication.**
Following the standard RCA cleaning process, a 4-inch 4H-SiC wafer with an epitaxial layer was prepared, featuring N-doping concentrations of $1\times10^{18}$ cm$^{-3}$ for the wafer and $5\times10^{14}$ cm$^{-3}$ for the epi-layer. The wafer and an oxidized silicon substrate were activated using 100 W $O_2$ plasma for 30 seconds. The two wafers were bonded together at room temperature under a pressure of 3000 N. To strengthen the bond, the structure was annealed at 800 °C for 6 hours before proceeding to the grinding stage. The grinding process was conducted in two phases: initially, a ten μm diamond slurry was used to grind the bonded wafer under a pressure of 600 N for around 10 hours, reducing the SiC layer to approximately 30 μm. Subsequently, the SiC layer was further ground down to 10 μm using a three μm diamond slurry. A chemical mechanical polishing (CMP) process was then employed to eliminate any surface damage caused by the grinding. The SiCOI wafer was diced into individual dies, with each die undergoing ICP dry etching to achieve the final thickness (200 nm). This etching was performed using 100 W RF and 1000 W ICP power at 10 mTorr. To introduce ensemble divacancy spins, we performed ion implantation using 30 keV carbon ions with a dose of $1.0\times10^{13}$ cm$^{-2}$. This was followed by annealing in a high vacuum at 900°C for 30 minutes to repair any remaining lattice damage.

The fabrication of photonic devices commenced with depositing a 30 nm Cr layer on the SiCOI membrane with ensemble spin. The micro-ring cavities with different diameters were then patterned onto an electron beam resist (ZEP 520A) utilizing electron-beam lithography with an overlay technique. Following resist development, the Cr mask was formed via ICP dry etching. The SiC membrane was etched to form the photonic devices guided by the Cr hard mask. The ICP etching was carried out utilizing 100 W RF and 1000 W ICP power at a pressure of 10 mTorr.

**Device characterization and spin control of ensemble spins.**
The fabricated micro-ring cavities were characterized using a home-built confocal microscope operating at a temperature of 5 K within a Montana Cryostation closed-cycle cryostat (CryoAdvance 50). A numerical aperture 0.65 near-infrared objective (Olympus, LCPLN50XIR) was employed, along with a single-mode fiber-coupled superconducting nanowire single-photon detector (PHOTEC). A 914 nm diode laser (MIL-III_914-300mW) was used to excite the PL4 divacancy spins, while a 50 μm copper wire served as the microwave antenna above the device. All experimental sequences were synchronized using a pulse blaster (Spincore, PBESR-PRO-500-PCI). In lifetime measurements, a picosecond 940 nm laser (NKT Photonics, PIL1-094-40FC) and a time-correlated single-photon counting system (ID1000) were utilized.

**Data availability**
Source data to generate figures and tables are available from the corresponding authors on reasonable request.

**Acknowledgments**

We acknowledge the support from the National Key R&D Program of China (Grant No. 2021YFA1400802, 2022YFA1404601, 2023YFB2806700), the National Natural Science Foundation of China (Grant No. 12304568, 11934012, 62293520, 62293522, 62293521, 12074400 and 62205363), the GuangDong Basic and Applied Basic Research Foundation (Grant No. 2022A1515110382), Shenzhen Fundamental research project (Grant No. JCYJ202412023000152, J20230807094408018), Guangdong Provincial Quantum Science Strategic Initiative (GDZX2403004, 2303001, GDZX2306002, GDZX2200001), Young Elite Scientists Sponsorship Program by CAST, New Cornerstone Science Foundation through the XPLORER PRIZE, Shanghai Science and Technology Innovation Action Plan Program (Grant No. 22JC1403300), CAS Project for Young Scientists in Basic Research (Grant No. YSBR-69), the Major Key Project of PCL, the Talent Program of Guangdong Province (Grant No. 2021CX02X465).

**Author contributions**
Y.Z. and Q.S. conceived the idea. A.Y. and X.O. prepared the SiC membrane. Q.L., T.B. and S.X. carried out the EBL lithography and Device fabrication. T.B. and Y.Z. built the setup and carried out the measurements. T.B., Y.Z., and Q.S. performed the simulations. Y.Z., T.B., and Q.S. wrote the manuscript. All authors contributed to analyzing the data and commenting on the manuscript.

**Competing interests**

The authors declare no competing interests.

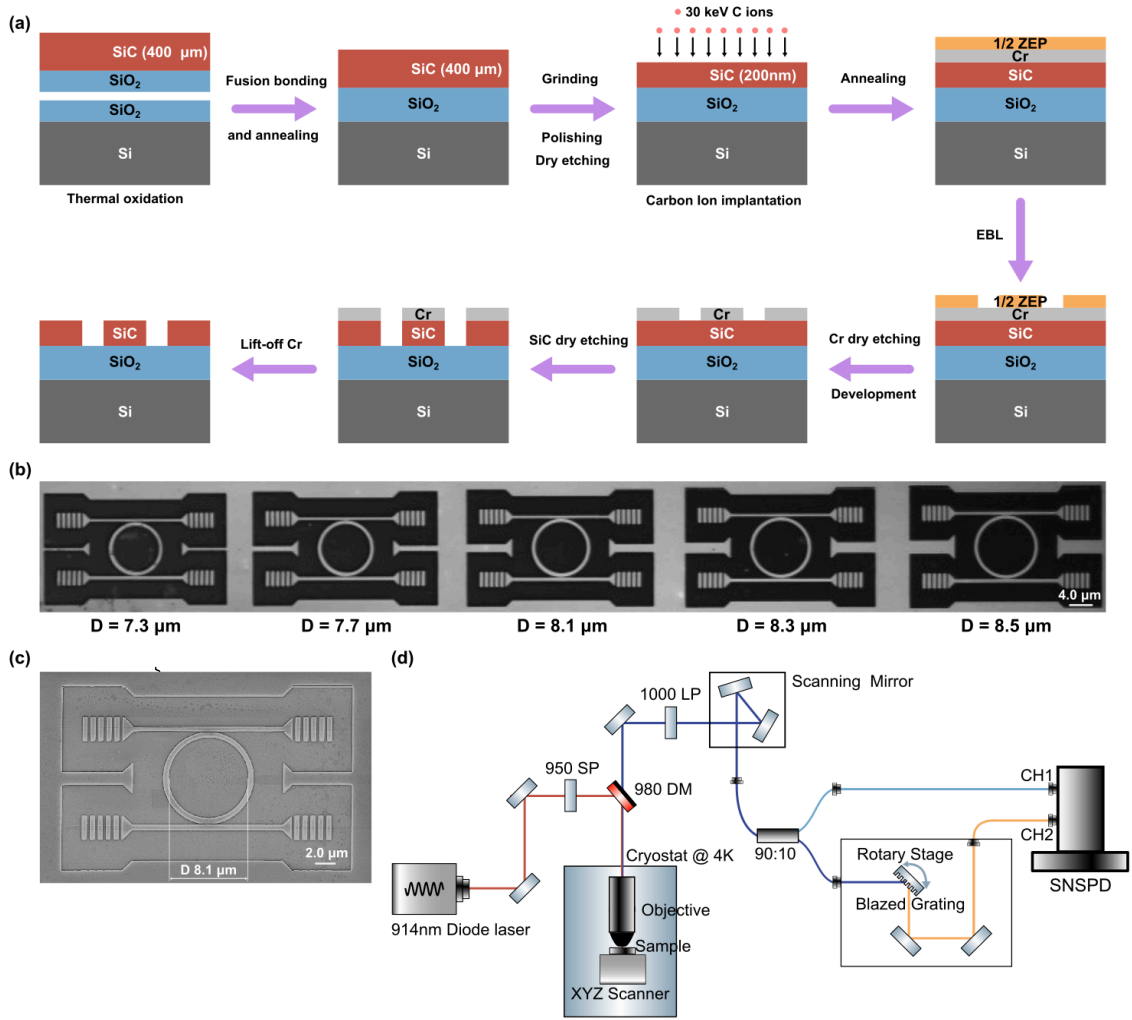

**Figure 1. PL4 divacancy spins integrated micro-ring cavities on 4H-SiCOI. (a)**. A 4H-SiC wafer with an epitaxial layer and a Si wafer were oxidized to form a thin layer of SiO2 on the surface. The two wafers were then fusion-bonded through annealing to enhance the bonding strength. The 400 μm thick SiC membrane was thinned to a target thickness of 200 nm through successive grinding, mechanical polishing, and ICP dry etching. Divacancy spins were introduced into the membrane via carbon ion implantation followed by high-vacuum annealing. The fabrication of photonic devices began with the deposition of a 30 nm Cr layer onto the SiCOI substrate containing the ensemble spins. Micro-ring cavities of varying diameters were patterned onto an electron beam resist (ZEP 520A) using electron beam lithography (EBL). After the resist was developed, a Cr mask was created through ICP dry etching, which was used to etch the designed photonic devices into the SiC substrate. Finally, the Cr mask was lifted off, and the devices were fabricated. **(b)**. Optical image of fabricated miro-ring cavities with different diameters from 7.3 *μm to* 8.5 *μm*. **(c)**. Scanning electron microscopy (SEM) image of the micro-ring cavity with a diameter of 8.1 *μm.* **(d)**. Optical setups used for device characterization. A custom-built 4K microscopy system collects the fluorescence from PL4

divacancy spins. The emission is directed to a 90:10 beam splitter for either optically detected magnetic resonance (ODMR) and spin control (10%, CH1 of SNSPD) or PL spectrum and lifetime measurements (90%). A blazed grating with a rotary stage is used in the PL spectrum measurements.

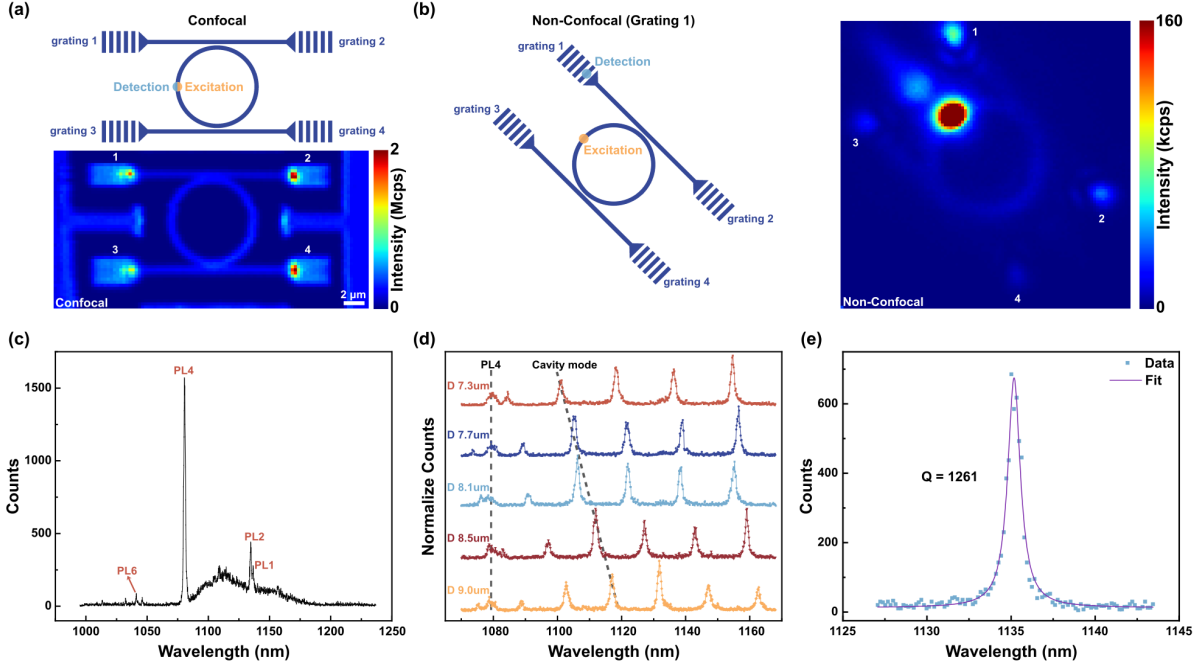

**Figure 2.** Optical characterization of the micro-ring cavities at 4K. **(a)**. The confocal scan map of the micro-ring cavity at 4K shows that the micro-ring and coupling waveguides, equipped with four grating couplers, align precisely with the designed geometry. **(b)**. To collect the emission from the grating coupler 1, the excitation spot is maintained at the ring while the collection angle is varied using a scanning mirror in the collection arm. Four distinct emission spots from the grating couplers are clearly visible and marked, confirming that the divacancy PL can be coupled to the waveguide and emitted from the side grating couplers. **(c)**. The full spectrum of divacancy PL at 0.2 mW excitation power and 4K is shown. The dominant peak is the zero-phonon line (ZPL) from PL4, while ZPLs from PL1, PL2, and PL6 are also observable, though with much lower intensity. **(d)**. The PL spectrum acquired from grating 1 in Figure B shows different micro-ring cavities with varying diameters under 0.4 mW excitation power. Multiple cavity modes are visible in the phonon sidebands with a free-spectral range (FSR) $14.8\pm0.07$ nm to $17.2\pm0.02$ nm. By altering the diameters, the cavity modes exhibit progressive wavelength shifts, as indicated by the dashed line. The ZPL from PL4 remains unchanged and weak, denoted by a dashed line, indicating it's not coupled with the cavity modes. **(e)**. The measured cavity mode coupled with PL4 luminescence shows a quality factor (Q) of $1261\pm39$. A Lorentzian fit to the raw data is displayed in red.

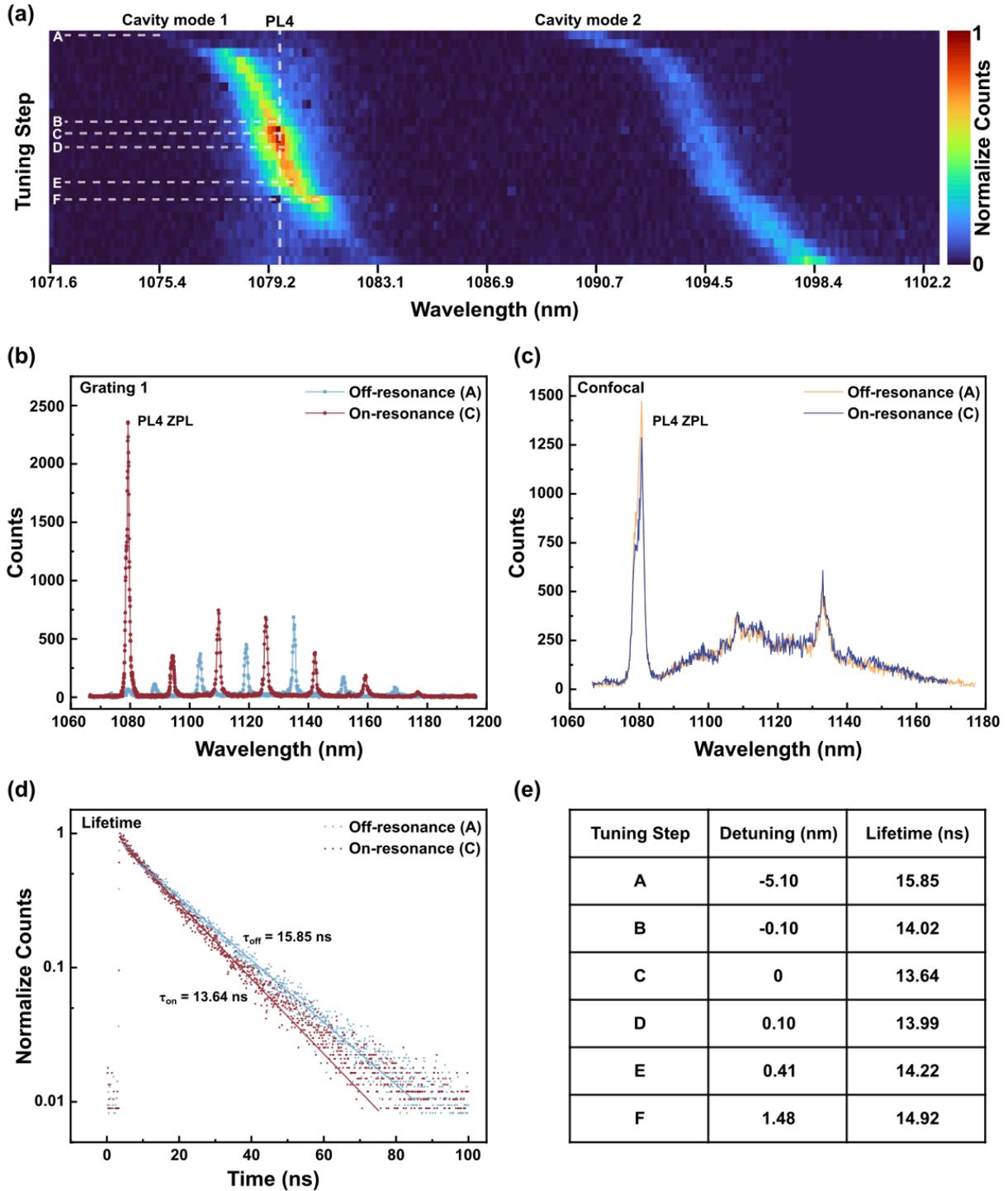

**Figure 3. Continuous cavity mode tunning via gas condensation**. **(a)**. The intensity map shows the redshift of the cavity mode as nitrogen gas is continuously injected, causing the cavity mode to intersect the fixed ZPL of PL4 and resulting in significant emission enhancement. Points A-F represent specific stages within the continuous gas condensation process. **(b)**. PL spectrum comparison between off-resonance (Point A) and on-resonance (Point D) under 0.2 mW excitation, shows a 36-fold increase in ZPL intensity. **(c)**. Confocal PL spectra for on- and off-resonance conditions show similar intensity, confirming that the emission enhancement is well confined within the cavity and the waveguide, finally emitted

from the output grating coupler. **(d)**. Lifetime measurements of the ZPL fitted with a single exponential function, indicating a reduction in lifetime from $15.85\pm0.09$ ns off-resonance to $13.64\pm0.07$ ns on-resonance. **(e)**. Summary of lifetimes at points A-F, demonstrating that the closer the cavity mode is to the ZPL, the more pronounced the reduction in lifetime.

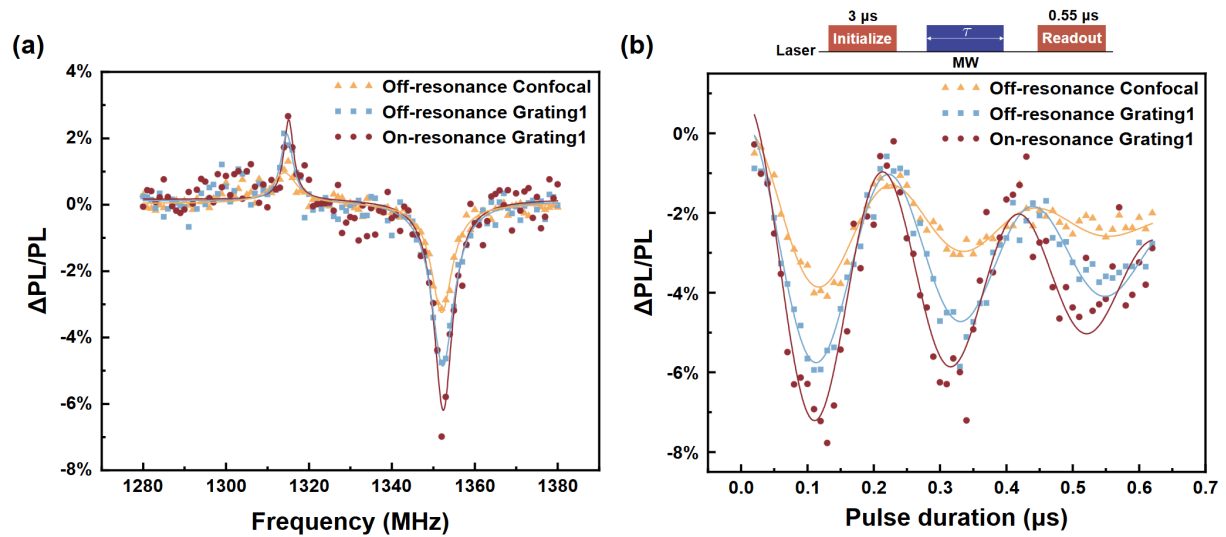

**Figure 4. ODMR and coherent control of the PL4 spin in micro-ring cavities. (a).** Three ODMR measurements were obtained in (i) off-resonance from confocal scan (ii) off-resonance from grating 1, and (iii) on-resonance from grating 1. All spectra reveal two peaks at $1315.1\pm0.3$MHz and $1352.4\pm0.2$MHz. From Lorentzian fitting, the ODMR contrast is extracted to be 3.2%, 4.8%, and 6.2%, respectively. On-resonance from the Grating 1 case showed a twofold enhancement in ODMR contrast, highlighting the significant contribution of the ZPL filtering and boosting effect of the microcavity. **(b).** Rabi oscillations of the PL4 divacancy spin, demonstrating coherent control of PL4 spins both coupled and uncoupled with the micro-ring cavity, in alignment with the observed ODMR contrast. The laser and microwave pulse sequences utilized for measuring the Rabi oscillations are illustrated.